# Hawking Radiation of Warped Anti de Sitter and Rotating Hairy Black Holes with Scalar Hair

Huriye Gürsel

Submitted to the
Institute of Graduate Studies and Research
in partial fulfillment of the requirements for the degree of

Master of Science
in
Physics

Eastern Mediterranean University
September 2015
Gazimağusa, North Cyprus

Approval of the Institute of Graduate Studies and Research

                                                                         Prof. Dr. Serhan Çiftçioğlu
                                                                               Acting Director

I certify that this thesis satisfies the requirements as a thesis for the degree of Master of Science in Physics.

                                                                     Prof. Dr. Mustafa Halilsoy
                                                          Chair, Department of Physics

We certify that we have read this thesis and that in our opinion it is fully adequate in scope and quality as a thesis for the degree of Master of Science in Physics.

                                                               Assoc. Prof. Dr. İzzet Sakallı
                                                                      Supervisor

                                                                                     Examining Committee

1. Prof. Dr. Özay Gürtuğ                                                 __________________________

2. Prof. Dr. Mustafa Halilsoy                                       __________________________

3. Assoc. Prof. Dr. İzzet Sakallı                                    __________________________

# ABSTRACT

This thesis mainly focuses on the Hawking radiation (HR) evacuating from the surface of the objects that have earned a reputation as the most extraordinary objects existing so far; the black holes (BHs). Throughout this study, quantum tunneling (QT) process serves as the model for the HR of scalar, vector and Dirac particles. The scalar and Dirac particles are anticipated to be tunneling through the horizon of rotating scalar hairy black holes (RHSBHs); whilst the vector particles are associated with a rotating warped anti de-Sitter black hole (WAdS$_3$BH) embedded in a (2+1) dimensional fabric. It is no coincidence that for all three cases; the standard HT expression is derived. Additionally, the engagement of conformal field theory (CFT) with anti de-Sitter (AdS) space presents itself to the reader and the methodologies of Klein-Gordon equation (KGE), Dirac equation and Proca equations (PEs) are introduced. For all three cases, Hamilton-Jacobi (HJ) approach is used, together with Wentzel-Kramers-Brillouin (WKB) Approximation. Finally, the BH structures are compared with thermodynamics to derive HT with the aid of Boltzmann factor.

# ÖZ

Bu tez, bugüne kadar karşımıza çıkan en gizemli maddeler olarak ün salmış kara deliklerden yayilan Hawking radyasyonu (HR) hakkindadir. Bu tezde, skaler, vektörel ve Dirac parçacıklarından oluşan HR'u, kuantum tünelleme (QT) esas alınarak incelenmiştir. Skaler ve Dirac parçacıklarının, dönen saçlı skaler kara deliklerden (RHSBHs); vektör parçacıklarının ise (2+1) boyutlu uzay zamana ait dönen eğrilmiş anti de-Sitter kara deliklerden (WAdS$_3$BHs) yayıldığı varsayılmıştır. Her üç durum için de hesaplanan Hawking sıcaklığı (HT) standarttır. Aynı zamanda, conformal field teorisi'nin (CFT), anti de-Sitter (AdS) uzayı ile olan ilişkisi kendini okuyucuya takdim etmiş; Klein Gordon denklemi (KGE), Dirac denklemi ve Proca denklemleri (PEs) tanıtılmıştır. Hamilton Jacobi (HJ) denklemi, Wentzel-Kramers-Brillouin (WKB) yaklaştırması ile birlikte kullanılmıştır. Son olarak, kara delik yapıları termodinamik kanunları ile karşılaştırılmış ve Boltmann faktörü kullanılarak Hawking sıcaklığı hesaplanmıştır.

**Anahtar Kelimeler**: Hawking radyasyonu, kuantum tünellemesi, Klein-Gordon denklemi, Dirac denklemi, Proca denklemi, skaler saç, Hamilton-Jacobi metodu, kuantum yerçekimi.

# DEDICATION

Dedicated to my grandmother, Huriye Hüdaoğlu...

# ACKNOWLEDGEMENT

Firstly, I would like to express my sincere gratitude to my respected supervisor Assoc. Prof. Dr. Izzet Sakalli for his endless support, motivation and guidance. Without his supervision and remarkable knowledge, this thesis would not have been possible to be completed. His guidance helped me in all stages of this research and made this a thoughtful and fulfilling voyage. I am extremely thankful and indebted to him for experiencing his valuable encouragement throughout these two years. Besides my supervisor, my genuine thanks also goes to all of the Department faculty members; the esteemed chairman Prof. Dr. Mustafa Halilsoy, the instructors and Mr. Resat Akoglu and Mrs. Cilem Aydıntan for their help and support over the past two years. I would also like to express my appreciation to all my friends in the faculty, especially to Danial for the inspiring discussions and his valuable support and to Ali for his encouragement and insightful comments and to Sara for her support.

In particular, I am grateful to my dear friends Sina and Iman for spending countless hours on the format check, proof reading and their endless support alongside. Without their support, it would have been extremely hard to finish up the thesis on time. I would like to add my special gratitudes for my friends Mobina and Elnaz for showing their worthful support, my old friend Hidayet for generously sharing her time and the stimulating discussions going on for sleepless nights. I am also extremely grateful to Harry Savy who supported me throughout this venture with patience and care and provided me with proof reading.

Last but certainly not the least, I would like to take this opportunity to thank my parents Cebriye and Ibrahim Gursel, my brother Ahmet and my whole family for supporting me spiritually throughout my life. The unceasing encouragement, support and attention I have received from them have definitely enabled me to carry on my further studies. Furthermore, I truly appreciate my father for checking my grammar in detail with an infinite patience and for sharing his precious knowledge with me.

# TABLE OF CONTENTS

# LIST OF TABLES

# LIST OF FIGURES

# Chapter 1

# INTRODUCTION

If one dwells on the past, then he robs the present. However, if one ignores the past, he may rob the future. Have you ever wondered to comprehend what actually is happening behind the scenes in life? In point of fact, physics cardinally emerges from curiosity. The more one thinks about how the universe was created, the deeper s/he moves into the field of physics. For decades, physicists have been carrying out experimental and theoretical research for being able to obtain more information regarding the universe that we live in. This thesis has also eventuated from certain questions asked due to the inspiring curiosity. If one is able to squeeze the rules of nature in the equations provided by physics and mathematics, does there exist any place in the universe where these laws as we know them are not valid anymore? Can information ever be lost? Before the creation of galaxies, were there any other objects that were already in existence? All these questions divert us to the mysterious world of BHs.

As can be noticed from the title, we will be covering HR sparkling from a BH and we will be dealing with several methodologies in order for deriving the surface temperature associated with this radiation process, more widely known as the HT. Since the temperature is directly linked to the BHs, it is legitimate to first delve into the scope of BHs and the first question popping into our minds should be; 'what is a BH?'

BHs can be thought as hazardous deadfalls in which nothing can survive. They are formed as a consequence of stars bursting out loudly and violently, which marks them as awe-inspiring objects. What make them different from the rest of the universe are the enormously intense gravitational effects that they experience which cause laws of nature to be non-valid. There exists prevailing reasons to believe that BHs are the most mysterious objects we have come across with so far, hiding a lot of secrets waiting to be revealed regarding the past, present and future of our universe. The more we can find out about them, the more understanding we will gain about the nature of our universe.

The two distinctive constituents of a BH are its boundary and the singularity at the center, which is infinitely small and dense. Clearly, the singularity is the point that we are willing to stay away from, however it is worthy to point out that it should be the boundary that should concern us the most. Not because there exists a prospect of danger on the boundary, but due to the conjecture that once a particle passes the point of no return, or more commonly known as the 'event horizon', nothing will save it from being trapped inside and it will continue falling towards the singularity.

Having made mention of the structure of BHs, we may now move on to the question of how we can describe a BH mathematically. It is intuitive to recognize that each BH is represented by a characteristic metric of its own. Depending on the structure that it possesses, one can be equipped with the hallmarks of the BH concerned. Even though the metric of a BH seems to be considerably trivial, its importance should not be taken for granted. It plays a crucial role in the theory of BHs, not only because it obscures latent information regarding the BH's geometry, but also for enabling a plausible explanation of what 'gravity' means in the language of general relativity (GR). Rather than conceptualizing gravity as a type of force, one may refer to it as being a natural outcome of the curvature of spacetime. This idea offered itself as an outcome of a very critical question in Albert Einstein's mind. He kept on asking himself whether the gravity should be treated as a type of force or not. To be truthful, this was a basic concern for why Newton's interpretation of gravity was not adequate once light was tackled. Based on these reasons, Einstein developed his equations which will be studied in detail later on in this chapter.

The substantial punchline in Einstein's approach was that the concept of gravity in GR had to agree with Newton's formalism on classical level. Therefore, along the investigation of BHs, one should definitely give an enormous amount of credits to Albert Einstein's hard work on the theory of gravity.

Before we get around to the line element describing the structure of a BH, it is ineluctable to ask why it is of great importance. Metrics can be computed by taking into account the main constituents of our universe; matter and energy. Just like the DNA of a person, these two factors are the building stones determining the unique BH structures. Einstein had proposed his equations [1] in order for facilitating a mathematical expression to explain the physics of nature in one line. He managed to relate the curvature of spacetime to matter and energy, or in other words, the galactic dust surrounding us. His idea was outstanding and his equations had rather impressive solutions; the BHs. These famous equations of Einstein which is the fundament of GR read (Stephani, Kramer, MacCallum, Hoenselaers, & Herlt, 2003).

---

[1] It is known that Einstein's Field Equations are presented as a single equation. However, since it includes tensors of rank 4x4, I preferred to name them as equation's' throughout this thesis. Once they are examined explicitly, sixteen distinct equations arise.

$$R_{\mu\upsilon} - \frac{1}{2}Rg_{\mu\upsilon} + \Lambda g_{\mu\upsilon} = \frac{8\pi G}{c^4}T_{\mu\upsilon} \qquad (1.1)$$

where $R_{\mu\upsilon}$, $g_{\mu\upsilon}$ and $T_{\mu\upsilon}$ represent 4x4 matrices known as Ricci, metric and energy momentum stress tensors respectively; whereas $R$, $\Lambda$, $G$ and $c$ stand for Ricci scalar, cosmological constant, gravitational constant and speed of light. For simplicity, I will be using $c = G = 1$ throughout this thesis. Hence, Eq. (1.1) can be rewritten as

$$R_{\mu\upsilon} - \frac{1}{2}Rg_{\mu\upsilon} + \Lambda g_{\mu\upsilon} = 8\pi T_{\mu\upsilon} \qquad (1.2)$$

In some resources like (Carroll, 2004; Hartle, 2002), the term including the cosmological constant is ignored, since its value is rather small. However, its importance should not be looked upon lightly. From Eq. (1.2), it is quite clear that all the three terms at the left hand side of the equation describe the geometrical shape of spacetime to be considered. If we were to determine from which term to start our investigation on Einstein's equations, the metric tensor would be the most logical one to be chosen. The reason for this is because once the metric of the BH is known, the metric tensor is rather simple to be recognized. Consider a metric of the form (Carroll, 2001)

$$ds^2 = -dt^2 + dr^2 + r^2 d\theta^2 + r^2 \sin^2\theta d\phi^2, \qquad (1.3)$$

which describes flat spacetime with the time coordinate *t* and spatial spherical polar coordinates $(r, \theta, \phi)$. If we were willing to find the metric tensor belonging to Eq. (1.3), we would require the relation

$$ds^2 = \sum_{\mu\upsilon} \frac{\partial x^\mu}{\partial \varepsilon^\gamma} \frac{\partial x^\upsilon}{\partial \varepsilon^\kappa} \delta_{\mu\upsilon} dx^\gamma dx^\kappa, \qquad (1.4)$$

where the coefficient of $dx^\gamma dx^\kappa$ generates the elements of the metric tensor. For information regarding the computation of the Ricci tensor and Ricci scalar from the metric tensor, please assign (Einstein, 1952).

After all, why have we introduced Einstein's field equations and what is the relationship between these equations and BHs? When one can successfully solve these equations, the solutions obtained will be equivalent to the BHs. Furthermore, with the aid of these equations, one of the greatest mysteries in the physics world was unlocked: the reason of why we are able to stay still on the floor, sit down on an armchair and so on so forth. Einstein's equations victoriously implied that we are being forced by the curvature of spacetime to be pushed downwards. This opened a brand new chapter in physics and gave birth to a fascinating approach towards the concept of gravity.

It is worthwhile to mention that in curved spacetime, everything will be obliged to follow geodesics throughout their motion. The geodesic equation can be introduced as (Hobson, Efstathiou, & Lasenby, 2006)

$$\frac{d^2 x^\mu}{d\lambda^2} + \Gamma^\mu_{\rho\sigma} \frac{dx^\rho}{d\lambda} \frac{dx^\sigma}{d\lambda} = 0 \qquad (1.5)$$

where $\Gamma^\mu_{\rho\sigma}$ is known as the Christoffel symbol playing a core rothe Ricci tensorlculation of Ricci tensor and scalar curvature. The patterns followed, obeying Eq. (1.5), appear to fit the criteria supporting the hypothesis of particles trying to always travel in straight lines on a flat space. In the case where we consider curved surfaces, the geodesic equation is actually equivalent to Newton's laws of motion in classical mechanics while every day objects' motion in ordinary space is discussed.

When planets, galaxies, or more explicitly, the spherical objects existing in our universe are considered, the metric to be processed should be the Schwarzschild metric which experiences spherical symmetry. However, in this thesis, we are not interested in this metric, hence for further details, one may check (Mei, 2006).

The thesis is organized into five chapters. Chapter 2 provides a brief review of the thermodynamic properties of BHs. The motivation for this chapter arises from the outstanding similarities between black body radiation covered in statistical physics and HR phenomenon experienced on the surface of a BH. The chapter ends by providing a framework for the issue of HT evaluation with reference to the QT process. Throughout the subsequent chapters, I will be basing our findings on the QT process; therefore this chapter documents the importance of getting the logic behind this technique.

In Chapter 3, it is attempted to offer a theoretical foundation for the concept of evaluation of HT for RHSBHs by using two different approaches. Firstly, the particles which are tunneling through the horizon are postulated to be scalar and KGE is applied. This section is then followed by the replacement of these particles by Dirac particles and the final HT result is demonstrated via using Dirac equation.

Unlike previous studies, Chapter 4 explicitly takes into account applying PEs to find the HT of a $WAdS_3BH$. During this chapter, it is intended to make sense of and provide insight into AdS space and examine the engagement between CFT and BHs in $AdS_3$ space. Finally, the conclusions drawn are demonstrated in Chapter 5.

# Chapter 2

# BHS AND THERMAL PHYSICS

## 2.1 Thermodynamic Properties of BHs

According to the laws of thermodynamics, heat energy can be transferred from one place to another by conduction, convection and radiation (Young & Freedman, 2008). Radiation is a type of process where a medium to travel in is not required. All bodies emit electromagnetic (em) radiation and the temperature of a body is the factor determining the peak wavelength of this radiation (Resnick, Halliday, & Walker, 1988). Hawking and Bekenstein stated that BHs behave as black bodies, or in other words, they are bodies at a constant temperature that absorb and re-emit all the em radiation falling on them (John D. Cutnell, 2013).

I am inflicted with an intense faith of approaching concepts which appear to be tremendously complicated in a version where it is reduced to its bare bones. Whenever one is in need of learning something new in any stage of her/his life, regardless of the area it belongs to (for instance; a baby trying to figure out how to put her/his first step into practice, attempting to pick up how to play a relatively hard piano piece, seeking for new baking recipes, grasping new mathematical concepts, solving criminal cases, diagnosing the type of disease a patient is suffering from and so on), the first step is to create a perspective of our own and inspect the situation in the most simplified version. Hence, I would like to start by a naïve but rather appealing picturesque way of defining entropy, temperature and heat energy.

Suppose that a very considerate man, let us name him as John, with a gentle heart moves into a village with the aim of increasing the total number of people smiling in that village. John's purpose of the visit is to record the total number of smiles he observes per day and generate a rise in this number. To achieve his goal, he builds a farm of his own and starts producing dairy products. Whenever he is satisfied with the amount of products he has obtained, he starts visiting each house one by one and distributes the daily products free of charge. The more smiles he observes per day, the happier he gets. We can apply this scenario to the physics world in order to get a clear picture of the definitions of the term 'temperature', 'entropy' and 'heat energy'. As you might have guessed, the dairy products refer to the heat energy being transferred from one place to another, temperature is expressed by the eagerness of John to give the products to the peasants and lastly, the total number of smiles recorded per day (which is increasing as time passes by) represents the entropy. The entropy of a BH is the extent of molecular disorder and one can signify it as being equivalent to the quantum microstate degeneracy of the near horizon 2D CFT (Strorninger & Vafa, 1996). While the meaning of entropy is considered, one might feel like it is dependent on the volume to be taken into account. However, Hawking (S. W. Hawking, 1975) and Bekenstein (J.D Bekenstein, 1973) found out that this is not the case. They stated that entropy actually depends on the surface area, not the volume, which can mathematically be presented as

$$S_{BH} = \frac{A}{4}, \tag{2.1}$$

in which $A$ designates the area of the concerned BH's surface; or in other words, the event horizon area. One can go back to thermodynamics and have a look at the first law of it. It will not be of amazement to spot this relation obeys this law and can be derived by making the assumption that the temperature associated with the BH can be inscribed as

$$T_H = \frac{\kappa \hbar}{2\pi},\qquad(2.2)$$

where $\kappa$ stands for the surface gravity term. Eq. (2.2) is known as the HT which is the main focus of this thesis. A selection of approaches will be liberated for derivation of this manifestation during the incoming chapters. For further reading regarding the laws of BH thermodynamics, please perceive (Bardeen, Carter, & Hawking, 1973).

## 2.2 Exploring HR

In 1973, Stephen Hawking expected the unexpected and claimed that BHs are actually not black; they emit radiation from, let say a thin layer near their surface, which is now known as the HR.

Prior to Hawking's discovery, BHs were thought to be fully black. What made his discovery so enterprising was that it brought forth a linkage between thermodynamics and gravity. Going into BH's thermodynamic properties carefully can lead us to a precise explanation of what gravity actually is. Thus, let us dig into Hawking's remarkable theory: the HR.

Since Hawking proposed his theory, many explanatory approaches aroused with the aim of finding alternative ways of explaining the HR process. Throughout this thesis; I will be focusing on the particle-antiparticle approach, which is just one of the many promising hypothetical explanations behind HR.

From classical point of view, a BH is expected to only absorb radiation. However, it is more convenient to approach BHs from the semi-classical perspective. A particle – antiparticle pair is alleged to be produced near the horizon due to vacuum fluctuations. In other words, the boundary of the BH can experience both emission and absorption, depending on whether we have a particle or an antiparticle tunneling through the horizon. Our task is to find expressions for the probabilities of these tunneling processes to occur and this is the point where QT comes into stage. Thus, let us move on by scrutinizing this technique.

**2.2.1 HR as a QT Process**

QT is a semi-classical approach carried out to achieve the evaluation of HT at the surface of a BH. The key point is whether we choose to start with the assumption of these particles to be forming inside or outside of the BH. For each case, the QT picture alters slightly due to the direction of tunneling (Akhmedov, Akhmedova, & Singleton, 2006). It is relatively easier to picturize particle creation taking place outside the BH, since the antiparticle will easily cross the boundary (without even being aware of it) and we will not be facing any difficulties. However, for the case where the creation of a particle – antiparticle pair is assumed to be taking place inside the BH, the tunneling is expected to happen from inside to outside. This generates the obligation of the particle crossing the event horizon and leaving the BH! At this point, you must be noticing a logical inconsistency between how I had defined BHs in Chapter 1 and what I have just stated. Previously, I had written that a BH can be described as a type of jail full of danger, not letting anything (even light!) escape from it, once the horizon is passed. On the contrary, QT requires one member of the particle – antiparticle pair to escape the BH, if the vacuum fluctuations had caused them to be formed inside the BH. How can we resolve this tricky contradiction? The answer is hidden in quantum fluctuations and I would like to further investigate this for two different cases. In case 1, the antiparticle will be moving towards the BH, whereas in case 2, it will be the particle tunneling out of the BH. The reason why we have limited ourselves to these cases is for if the other possibilities were considered, we would be obtaining an extra factor of two in our final HT result. To prevent this problem from arising and due to energy conservation, the two cases below will be the only ones that we will work on.

- Case 1

Assume that a particle − antiparticle pair is created just outside the event horizon as shown in Figure 1. One may presume the antiparticle to be travelling towards the BH, while the particle is moving away from it. The antiparticle will be passing through the horizon into the BH without any difficulty, nothing unusual about that. The moment when the antiparticle passes the point of no return, as its name suggests, it will not be able to escape from the BH ever again. The particle moving away is what we name as the 'HR' and consequently, the BH, which now does not seem to be fully black, radiates off a radiation of all kinds.

In particle physics, antiparticles are treated as moving backwards in time; hence they are drawn from right to left in Feynman diagrams. Due to this reason, we must keep in mind that the antiparticle entering the BH has negative energy and it will cause the BH to reduce in size as it radiates.

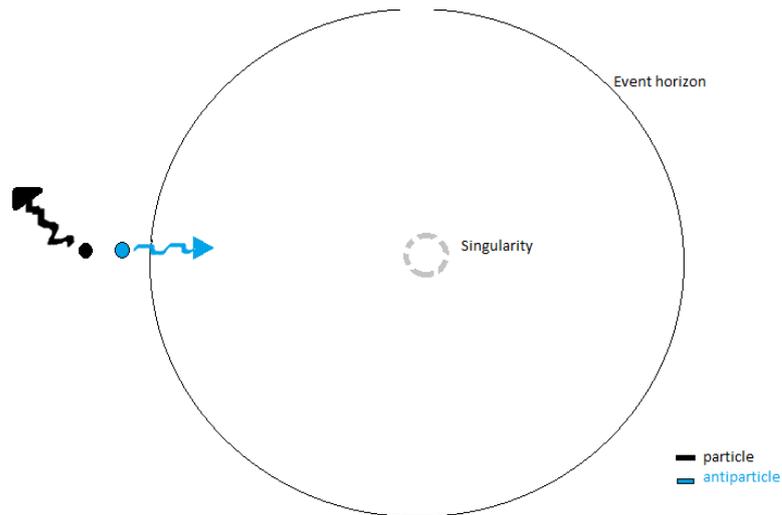

Figure 1. QT for the case when vacuum fluctuations create particle – antiparticle pairs just outside the event horizon.

- Case 2

Let us now change our starting point. What happens if we consider particle – antiparticle pairs to be created just inside the event horizon? This time, the particle can be picturized to tunnel out of the horizon (which is an event that would classically be prohibited), whilst the antiparticle starts its journey towards singularity. As for case 1, the mass of the BH will again be decreasing during the radiation process which is nothing but the particle tunneling outwards. One may check Figure 2 for a simple sketch of this case.

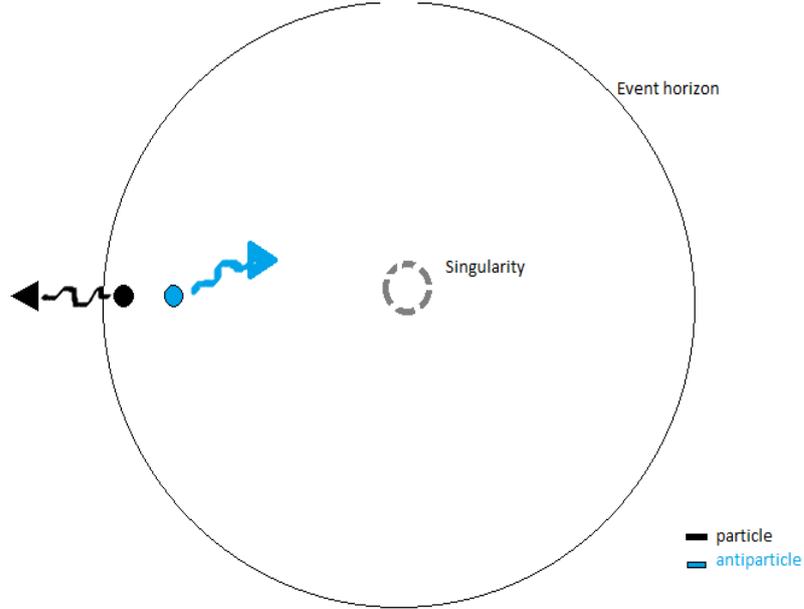

Figure 2. QT for the case when vacuum fluctuations create particle – antiparticle pairs just inside the event horizon.

### 2.2.2 Complex Contour Integral Approach

Prior to explaining the complex contour method, it is important to specify the metric to be used during our calculations, since it conceals marvelous information regarding the structure of the BH. In Chapter 3, we will be studying the line element

$$ds^2 = N(r)^2 dt^2 - \frac{dr^2}{N(r)^2} - r^2 \left(d\phi + N^\phi(r)dt\right)^2 \tag{2.3}$$

whilst in chapter 4

$$ds^2 = -N^2 dt^2 + \frac{\ell^4}{4R^2 N^2} dr^2 + \ell^2 R^2 \left(d\phi + N^\phi dt\right)^2 \tag{2.4}$$

will be used. Hence, I would like to explicate how to use complex contour analysis by taking

$$ds^2 = A(r)dt^2 - \frac{dr^2}{A(r)} \tag{2.5}$$

and the scalar fields as basis. Even though metric (2.5) is a naive version of what we will be dealing with throughout the incoming chapters, understanding how this mechanism works for the simplest case carries a tremendous importance in grasping the procedure. Once this version is understood, it can easily be applied to more complicated situations.

It is essential to note that the function $A(r)$ must have a non-zero radial derivative which is also finite at the boundary of the BH at $r = r_H$. Furthermore, Taylor expansion can be applied to this function for the points close to the event horizon, which permits us to denote (Arfken, 2013)

$$A(r) = A(r_H) + \frac{dA(r_H)}{dr}(r - r_H) + \mathrm{O}\left[(r - r_H)^2\right] = \frac{dA(r_H)}{dr}(r - r_H) + \mathrm{O}\left[(r - r_H)^2\right] \quad (2.6)$$

since $A(r_H) = 0$. This expression is beneficial, as complex contour integral analysis necessitates derivatives of the concerned functions. Having declared expressing our arbitrary function in terms of the radius of the horizon, let us now continue by witnessing how Eq. (2.6) will be of use.

Scalar particles obey the relation (Grössing, 2002; Izzet Sakalli, 2012)

$$\partial_\mu \left(\sqrt{-g}\, g^{\mu\upsilon} \partial_\upsilon \Psi_s\right) + \frac{m_s^2}{\hbar^2} \sqrt{-g}\, \Psi_s = 0, \quad (2.7)$$

of which $m_s$ and $\Psi_s$ represent mass and wave function of the scalar particles and $\partial$ stands for the partial derivative. This relation is acknowledged as the KGE. The wave function can be replaced by the änsatz

$$\Psi_s = \exp\left[\frac{i}{\hbar} \mathrm{I}(r,t)\right], \quad (2.8)$$

in which the action $I(r,t)$ can be further expressed as

$$I(r,t) = I_0(r,t) + \hbar I_1(r,t) + \hbar^2 I_2(r,t) + \ldots \quad (2.9)$$

Substituting Eq. (2.8) with the expanded form of the action in Eq. (2.7) and applying WKB Approximation which is allowed to be used in the theory of BHs (Gecim & Sucu, 2013), one can signify

$$\left(\frac{\partial I_0}{\partial t}\right)^2 - A(r)^2 \left(\frac{\partial I_0}{\partial r}\right)^2 - m_s^2 A(r) = 0 \ , \qquad (2.10)$$

which is nothing but the HJ equation. Clearly, to be able to move on, we need to substitute the time and radial derivatives of our action.

Note that the lowest order action $I_0$ is composed of two parts: time-dependent and space-dependent. It can be written as

$$I_0 = -Et + \omega(r) + \mathrm{h}\phi + \aleph, \qquad (2.11)$$

where $E$ and $h$ respectively represent energy and angular momentum and $\aleph$ is a complex constant. Hence, the derivatives we require are

$$\frac{\partial I_0}{\partial t} = \frac{\partial(-Et)}{\partial t} = -E \qquad (2.12)$$

and

$$\frac{\partial I_0}{\partial r} = \frac{\partial \omega(r)}{\partial r} \ . \qquad (2.13)$$

Substituting Eqns. (2.12) and (2.13) into Eq. (2.10) brings about

$$(-E)^2 - A(r)^2 \left(\frac{\partial \omega}{\partial r}\right)^2 - m_s^2 A(r) = 0 \qquad (2.14)$$

which should be solved for the radial function $\omega(r)$. Rearranging for the term including the partial derivative of radial function and solving for $\omega(r)$ yields to

$$\omega(r) = \pm \int \frac{1}{A(r)} \sqrt{E^2 - m_s^2 A(r)} \qquad (2.15)$$

On the horizon, the second term will vanish and plugging Eq. (2.6) into Eq. (2.15), we can alternatively express Eq. (2.15) as

$$\omega(r) = \pm \int \frac{E}{A'(r_H)(r - r_H)}, \qquad (2.16)$$

where

$$A'(r_H) = \frac{dA(r_H)}{dr}. \qquad (2.17)$$

According to the residue theorem (Arfken, 2013), under the condition that $f(z)$ is an analytic function on our contour, the following relation holds.

$$\frac{1}{\pi i} \int \frac{f(z)dz}{z - z_0} = f(z_0), \qquad (2.18)$$

iff $z_0$ is an interior singular point. In our case, an isolated simple pole exists at $r = r_H$. So, Eq. (2.18) shall be modified as

$$\frac{1}{\pi i} \int \frac{f(r)dr}{r - r_H} = f(r_H). \qquad (2.19)$$

Once Eq. (2.19) is compared with Eq. (2.16), it can be observed that the function $f(r)$ should be asserted as

$$f(r) = \frac{E}{A'(r_H)}, \qquad (2.20)$$

which entails

$$\omega_\pm = \pm i\pi f(r_H) = \pm \frac{i\pi E}{A'(r_H)}. \qquad (2.21)$$

where the positive and negative signs indicate emitted and absorbed particles respectively.

What we actually seek for are the imaginary terms in the action, as the virtual pairs are created due to the singularities to be inspected here (I Sakalli & Mirekhtiary, 2013). The only complexity exists in the radial function and the constant $\aleph$, which signifies

$$\mathrm{Im}(I_0) = \mathrm{Im}(\omega_\pm) + \mathrm{Im}(\aleph). \qquad (2.22)$$

For the antiparticles being absorbed, the equation is in the form

$$\mathrm{Im}(I_0)\big|_{abs} = \mathrm{Im}(\omega_-) + \mathrm{Im}(\aleph) \qquad ; \qquad (2.23)$$

whereas for the emitted particles

$$\mathrm{Im}(I_0)\big|_{ems} = \mathrm{Im}(\omega_+) + \mathrm{Im}(\aleph) \qquad (2.24)$$

holds. From Eq. (2.21), one can noticeably read that

$$\omega_+ = \frac{i\pi E}{A'(r_H)} \qquad (2.25)$$

and

$$\omega_- = -\frac{i\pi E}{A'(r_H)}. \qquad (2.26)$$

Thus, we can assertively state the relation

$$\mathrm{Im}(\omega_-) = -\mathrm{Im}(\omega_+) \qquad (2.27)$$

This equation will have an imperative importance during the incoming equations so we will be coming back to it shortly.

Now, let us go back to the inspection of probabilities of emission and absorption, for they provide flourishing information in the process of derivation of HT. Semi-classically, the probability of the absorption will be one, as BHs act as black bodies and the probability can be established by taking the square of the wave function considered; in our case, Eq. (2.8). Hence,

$$P_{abs} = \exp(-\frac{2iI}{\hbar}) = 1 \qquad (2.28)$$

allowing us to write

$$-\frac{2iI}{\hbar} = 0, \qquad (2.29)$$

followed instantly by

$$\text{Im}(I)\big|_{abs} = 0 . \qquad (2.30)$$

Therefore, Eq. (2.23) becomes

$$\text{Im}(\omega_-) + \text{Im}(\aleph) = 0 \qquad (2.31)$$

denoting that

$$\text{Im}(\aleph) = -\text{Im}(\omega_-) . \qquad (2.32)$$

Going back to the emitted particles, one can now substitute Eqns. (2.27) and (2.32) into Eq. (2.24) and rewrite the imaginary part of the action of particles being tunneled out of the horizon as

$$\text{Im}(I_0)\big|_{ems} = 2\,\text{Im}(\omega_+) . \qquad (2.33)$$

To sum up, one can draw the conclusion that the probabilities for emission and absorption of the particles and antiparticles can respectively be exhibited as

$$P_{ems} = |\Psi_{ems}|^2 = \exp(-2\,\text{Im}\,\omega_+) \qquad (2.34)$$

and

$$P_{abs} = |\Psi_{abs}|^2 = \exp(-2\operatorname{Im}\omega_-)$$

(2.35)

Familiar as we are with the probability expressions, we can now creep forward and expose the overall tunneling rate as (Shankaranarayanan, Srinivasan, & Padmanabhan, 2001) (I Sakalli & Ovgun, 2015)

$$\Gamma \equiv \frac{P_{ems}}{P_{abs}} = \exp(-4\operatorname{Im}\omega_+). \tag{2.36}$$

$$\Gamma \equiv \exp\left[-\frac{4\pi E}{A'(r_H)}\right] \tag{2.37}$$

from which the HT reads

$$T_H = \frac{A'(r_H)}{4\pi}. \tag{2.38}$$

with the aid of the Boltzmann factor stating (Schroeder, 2000)

$$\Gamma \equiv \exp\left(-\frac{E}{T}\right). \tag{2.39}$$

The HT expression (2.38) will be shown to remain unchanged throughout the thesis.

# Chapter 3

# HR OF THE ROTATING BHS WITH SCALAR HAIR IN 3D

In the theory of BHs, it has been extensively believed that BHs have no hair (Falls & Litim, 2014), which is known as the 'No Hair Theorem'. What is the meaning of the statement 'hair' in this expression? If we are referring a BH to be 'bald', what we actually have in mind is that our concerned BH can be portrayed only by referring to its mass, angular momentum and charge. (Note that the charge can be electric charge; or in the case of the hypothetical magnetic charge's existence is ever experimentally verified, then it would imply magnetic or electric charge; or both.) Therefore, if the BH is not bald, there exists a supplementary characteristic which contributes to the depiction of the BH. In this case, the no hair theorem as stated by Bekenstein (Jacob D Bekenstein, 1995) will be violated for some specific occasions. Some examples of these occasions can be given as SU (N) Einstein-Yang-Mills theory where there subsists a negative cosmological constant and RHSBHs (Toubal, 2010); which is the case we will be exploring.

## 3.1 Fundamental Attributes of RHSBHs

The metric in which the geometrical properties of our concerned BH are encoded can be written as

$$ds^2 = N^2 dt^2 - \frac{dr^2}{N^2} - r^2 \left(d\phi + N^\phi dt\right)^2, \qquad (3.1)$$

where

$$N^2 = \frac{r^2}{36\ell^2}\left(J^2\ell^2 x^2 - 12M\ell^2 x + 36\right)$$

(3.2)

and

$$N^\phi = \frac{-J}{6}\left(\frac{3r+2c}{r^3}\right), \qquad (3.3)$$

in which $c$ and $J$ stand for the scalar charge and angular momentum respectively. For simplicity, let us use the substitution

$$\Im = \frac{3r+2c}{r^3}, \qquad (3.4)$$

which implies that Eq. (3.3) should take the form

$$N^\phi = -\frac{J\Im}{6}. \qquad (3.5)$$

It is also possible to express Eq. (3.2) in a more compact version as

$$N^2 = \frac{J^2 r^2}{36}\left[(\Im-\Im_1)(\Im-\Im_2)\right] \qquad (3.6)$$

in which $\Im_1$ and $\Im_2$ differ due to

$$\Im_i = \frac{6}{J^2}\left[M - (-1)^i \frac{1}{\ell}\sqrt{M^2\ell^2 - J^2}\right]. \qquad (3.7)$$

At the surface of the BH where $r = r_H$, the metric coefficient $N^2$ vanishes (Bojowald, 2012). Hence, one can state

$$\frac{J^2 r_H^2}{36}\left[(\Im-\Im_1)(\Im-\Im_2)\right] = 0, \qquad (3.8)$$

which implies that

$$(\Im-\Im_1)(\Im-\Im_2) = 0; \qquad (3.9)$$

since angular momentum is non-zero. Let us examine Eq. (3.9) in two parts.

<u>Part 1</u> - $(\mathfrak{I}-\mathfrak{I}_1)=0$

Substituting Eqns. (3.4) and (3.7) with $i=1$ into the first part of Eq. (3.9) enables us to write

$$\frac{3r_H+2c}{r_H^3}-\left\{\frac{6}{J^2}\left[M+\frac{1}{\ell}\sqrt{M^2\ell^2-J^2}\right]\right\}=0. \tag{3.10}$$

Now, one can equate the denominators of each term and present the equation

$$\frac{(3r_H+2c)J^2\ell-6M\ell r_H^3-6r_H^3\sqrt{M^2\ell^2-J^2}}{r_H^3 J^2 \ell}=0, \tag{3.11}$$

which clearly vanishes to

$$r_H^3-\frac{3J^2\ell r_H}{6\left(M\ell+\sqrt{M^2\ell^2-J^2}\right)}-\frac{2c^3 J^2\ell}{6\left(M\ell+\sqrt{M^2\ell^2-J^2}\right)}=0. \tag{3.12}$$

<u>Part 2</u> - $(\mathfrak{I}-\mathfrak{I}_2)=0$

The same procedure will now be applied for $i=2$. In this case, one gets

$$r_H^3-\frac{3J^2\ell r_H}{6\left(M\ell-\sqrt{M^2\ell^2-J^2}\right)}-\frac{2c^3 J^2\ell}{6\left(M\ell-\sqrt{M^2\ell^2-J^2}\right)}=0. \tag{3.13}$$

From Eq. (3.9), it is clear that both parts 1 and 2 should hold. Therefore, one can combine the two by the following compact statement.

$$r_{H(i)}^3+\tilde{\mathfrak{I}}_i r_{H(i)}+\tilde{c}_i=0 \tag{3.14}$$

where

$$\tilde{\mathfrak{I}}_i=-\frac{3}{\mathfrak{I}_i} \tag{3.15}$$

and

$$\tilde{c}_i = -\frac{2c}{\Im_i} . \qquad (3.16)$$

The resolution of this problem will require a more detailed analysis of Eq. (3.14). To be able to solve this equation, we first need to find the determinant, namely $b_i$, which was evaluated as

$$b_i = \frac{4\tilde{\Im}_i^3 + 27\tilde{c}_i^2}{108} . \qquad (3.17)$$

For further details of how to obtain the discriminant, please check (Cox, 2011).

The discriminant (3.17) can be rewritten in the form

$$b_i = c^6 \frac{(\Im_i - 1)}{\Im_i^3} , \qquad (3.18)$$

once (3.15) and (3.16) are placed in Eq. (3.17). The solutions shine through in in two different cases; for $b_i > 0$ and for $b_i < 0$. The positive roots lie at the very core of

$$r_{H(i)} = \frac{c}{\Im_i}\left[\left(\Im_i^2 + \sqrt{\Im_i^3(\Im_i - 1)}\right)^{1/3} + \left(\Im_i^2 - \sqrt{\Im_i^3(\Im_i - 1)}\right)^{1/3}\right] \qquad (3.19)$$

$$r_{H(i)} = \frac{2B}{\sqrt{x_i}}\cos\left[\frac{1}{3}\cos^{-1}\left(\sqrt{x_i}\right)\right] \qquad (3.20)$$

To see how these equations are derived, please refer to (Zou, Liu, Wang, & Xu, 2014).

Finally, it is substantial to make a record of the unique hallmarks of RHSBHs (Zou, Liu, Zhang, & Wang, 2014) which makes them distinguishable from other BH types. The mass of the RHSBH can be expressed as

$$M_{BH} = \frac{J^2\ell^2(2c+3r_+)^2 + 36r_+^6}{12\ell^2 r_+^3(2c+3r_+)} , \qquad (3.21)$$

whereas the HT and entropies read

$$T_H = \frac{\kappa \hbar}{2\pi} = \frac{(c+r_+)\hbar \left[36r_+ - J^2\ell^2(2c+3r_+)^2\right]}{24\pi\ell^2 r_+^5 (2c+3r_+)} \qquad (3.22)$$

and

$$S_{BH} = \frac{A_H}{4}\left(1 - \frac{1}{8}\phi^2(r_+)\right) = \frac{4\pi r_+^2}{c+r_+}, \qquad (3.23)$$

of which a potential of the form

$$\phi(r) = \pm\sqrt{\frac{8c}{c+r}} \qquad (3.24)$$

can be introduced.

The angular frequency $\Omega$ and the function $N^\phi$ at the event horizon can be denoted as

$$\Omega = -\frac{g_{t\phi}}{g_{\phi\phi}}\bigg|_{r=r_+} = -N^\phi(r_+) = \frac{(3r_+ + 2c)J}{6r_+^3}. \qquad (3.25)$$

## 3.2 QT of Scalar Particles from 3D RHSBHs

In quantum field theory (QFT) (Kaku, 1993), one can investigate fields in four different categories: scalar, vector, tensor and spinor fields. Quantization of scalar field gives rise to scalar particles[2], vector field to vector particles and so on. The categorization of fields is carried out according to the unique spin values that these particles possess. Our focus is to examine HT in the case when scalar particles are concerned, which requires spin value to be zero. By comparing the HT expression that will eventually be obtained at the end of this section with the result in section 3.3, we aim to argue that no matter what spin value the particles in a RHSBH have, the HT outcome must stay invariant.

---

[2] For further detailed information on derivation of particle creation from fields via quantization, please refer to (Zou, Liu, Wang, et al., 2014).

Since we are examining scalar field, it is convenient to use KGE

$$\partial_\mu \left(\sqrt{-g}\, g^{\mu\upsilon} \partial_\upsilon \Psi_s\right) + \frac{m_0^2}{\hbar^2}\sqrt{-g}\,\Psi_s = 0. \tag{3.26}$$

In component form, this equation can further be expressed as

$$\partial_0\left(rg^{00}\Psi_{s,0}\right)+\partial_1\left(rg^{11}\Psi_{s,1}\right)+\partial_2\left(rg^{22}\Psi_{s,2}\right)+\partial_0\left(rg^{02}\Psi_{s,2}\right)+\partial_2\left(rg^{20}\Psi_{s,0}\right)-\frac{m_s^2}{\hbar^2}r\Psi_s = 0. \tag{3.27}$$

Supposing the wave function for scalar particles as

$$\Psi_s = \exp\left(\frac{i}{\hbar}I + I_1 + O(\hbar)\right) \tag{3.28}$$

and substituting the elements of metric tensor, Eq.(3.27) turns into

$$\frac{1}{N^2}(\partial_0 I)^2 - N^2(\partial_1 I)^2 - \frac{N-r^2(N^\phi)^2}{N^2 r^2}(\partial_2 I)^2 - 2\frac{N^\phi}{N}(\partial_0 I \partial_2 I) - m_s^2 = 0. \tag{3.29}$$

We can carry on our voyaging by recalling the action term

$$I_0 = -Et + h\phi + \omega(r) + \aleph, \tag{3.30}$$

being composed of the energy $E$ and angular momentum $h$ of scalar particles and $\omega(r)$ is the radial function which is the term deserving the most attention for the rest of our calculations. The radial function reads

$$\omega_\pm(r) = \pm\int \frac{\sqrt{\tilde{E}^2 - N^2\left(\frac{h^2}{r^2} + m_s^2\right)}}{N}, \tag{3.31}$$

or shortly

$$\omega_\pm(r) = \pm\int \frac{\tilde{E}}{A'(r_+)(r-r_+)}; \tag{3.32}$$

where

$$A(r) = N^2(r),\tag{3.33}$$

and the modified energy is nothing but

$$\tilde{E} = E - \hbar\Omega_H.\tag{3.34}$$

As stated before, the compact notation for radial function of incoming and outgoing particles is

$$\omega_\pm = \pm i\pi \frac{\tilde{E}}{A'(r_+)}.\tag{3.35}$$

All in all, the final HT expression can be written as

$$T_H = \frac{\hbar(c+r_+)\left[36r_+ - J^2\ell^2(2c+3r_+)^2\right]}{24\pi\ell^2 r_+^5(2c+3r_+)}.\tag{3.36}$$

## 3.3 QT of Dirac Particles from 3D RHSBHs

In this section, the *vierbein* formalism, or in other words the *tetrad* formalism, will be used. Hence, before we start our profound exploration on the Dirac equation in three dimensions, it is beneficial to introduce the main highlights behind this formalism.

The vierbein field theory is a theory proposed by Albert Einstein in 1928 (Einstein, 1928a, 1928b) as an alternative recipe of gravitation. In other words, he asserted a new technique of expressing curved spacetime manifold. Even though the formalism possesses difficulties of its own, it can be explicated in a rather fuss-free way, which is what I will be aspiring after throughout this section.

The main constituents of the vierbein[3] formalism are Dirac matrices, spinorial affine connection and spin operator. While mentioning Dirac matrices, the key argument is to point out the necessity of modifying them in such a way that they obey the curved spacetime structure. It is not promising to apply the definitions that hold in Minkowski spacetime to Einstein's GR. To be more expositive, in tensor notation, one can write

$$\sigma^{\mu}_{DC} = e^{\mu}_{k}\sigma^{k}_{DF}, \qquad (3.37)$$

in which the subscript *'DC'* refers to Dirac matrices in curved spacetime. On the other hand, *'DF'* stands for Dirac matrices in flat spacetime. For further information, see (Dirac, 1981). It is worthy to record that the contravariant index $\mu$ runs from 0 to 2 and so does the Latin index $k$, however $k$ is a dummy index unlike $\mu$. Furthermore, Table 1 articulated below can be used as a detailed guideline for the explicit computation of these expressions.

Table 1. The linkage between Dirac matrices in flat and curved spacetimes

| Dirac Matrices under the influence of curvature ($\sigma^{\mu}_{DC}$) | Computation by using Dirac Matrices in flat spacetime ($\sigma^{k}_{DF}$) |
|---|---|
| $\sigma^{0}_{DC}$ | $e^{0}_{0}\sigma^{0}_{DF} + e^{0}_{1}\sigma^{1}_{DF} + e^{0}_{2}\sigma^{2}_{DF}$ |
| $\sigma^{1}_{DC}$ | $e^{1}_{0}\sigma^{0}_{DF} + e^{1}_{1}\sigma^{1}_{DF} + e^{1}_{2}\sigma^{2}_{DF}$ |
| $\sigma^{2}_{DC}$ | $e^{2}_{0}\sigma^{0}_{DF} + e^{2}_{1}\sigma^{1}_{DF} + e^{2}_{2}\sigma^{2}_{DF}$ |

Having grasped the Dirac matrices, it can be seen beyond any doubt that without knowing the values for Dirac matrices in flat spacetime, it is not possible to move

---

[3] The words 'vier' and 'bein' are both German words. Vier means four, whereas bein means leg. Thus, vierbein formalism can be thought as 'tetrad formalism'.

forward. Dirac matrices in Minkowski space can be denoted in terms of the well-recognized Pauli matrices as

$$\sigma_{DF}^k = \left(\sigma_P^3, i\sigma_P^1, i\sigma_P^2\right),  \tag{3.38}$$

where $\sigma_P^a$ indicates Pauli matrices and $a = \{1,2,3\}$. The Latin index '$a$' is not assigned with the value zero, for $\sigma_P^0$ may imply identity matrix in some resources. For further calculations, we will be requiring the Pauli matrix values for each component (Condon, 1929), so let us express it in a compact matrix form as follows

$$\sigma_k = \frac{\hbar}{2}\begin{pmatrix} \delta_{k3} & \delta_{k1} - i\delta_{k2} \\ \delta_{k1} + i\delta_{k2} & -\delta_{k3} \end{pmatrix}  \tag{3.39}$$

where the Kronecker delta has the property

$$\delta_{km} = \begin{cases} 0 \text{ if } k \neq m \\ 1 \text{ if } k = m \end{cases}  \tag{3.40}$$

and for simplicity one may assume $\hbar = 1$ during this section. As can be remarked from Eq. (3.39), the definition of Pauli matrix components is given in covariant form. However, during our calculations, we will be using contravariant Dirac matrices. Since Dirac matrices are in direct relation with Pauli matrices, it is advantageous to note that for Pauli matrices, the covariant and covariant notations take exactly the same value. To sum up, we do not need to apply any changes to Eq. (3.38).

The Dirac equation can be illustrated as (Sucu & Unal, 2007)

$$i\sigma_{DC}^\mu \left(\partial_\mu - \Gamma_\mu\right)\Psi_f = \frac{m_f}{\hbar}\Psi_f,  \tag{3.41}$$

where $\Psi_f$ and $m_f$ respectively represent the wave function and mass of fermions, $\hbar$ is the well-known reduced Planck constant, $\Gamma_\mu$ stands for the connection and $\sigma_{DC}^\mu$ implies Dirac matrices in curved spacetime as had already been mentioned. The term in brackets is nothing but the definition of covariant derivative in curved space. Without the correction term, we would still be dealing with partial derivatives only, which would not be sufficient since GR is taken into consideration.

In order for being able to work out the distinctive Dirac equation for a RHSBH, we need to rewrite the metric (3.1) in a way that the coefficients of *dt*, *dr*, $d\phi$ and the cross term becomes distinguishable. Ultimately, the line element (3.1) can be illustrated as

$$ds^2 = \left[N^2 - (N^\phi)^2 r^2\right]dt^2 - \frac{1}{N^2}dr^2 - r^2 d\phi^2 - 2N^\phi r^2 d\phi dt \qquad (3.42)$$

Why have we felt the necessity of expressing the metric in the form above? The reason for this is for we wish to evaluate the covariant and contravariant metric tensors directly from the line element. The procedure to be followed is quite straightforward.

From Eq. (3.42), the covariant and contravariant metric tensors can be interpreted as

$$g_{\mu\nu} = \begin{pmatrix} N^2 - r^2(N^\phi)^2 & 0 & -r^2 N^\phi \\ 0 & -\dfrac{1}{N^2} & 0 \\ -r^2 N^\phi & 0 & -r^2 \end{pmatrix} \qquad (3.43)$$

and

$$g^{\mu\nu} = \begin{pmatrix} \dfrac{1}{N^2} & 0 & -\dfrac{N^\phi}{N^2} \\ 0 & -N^2 & 0 \\ -\dfrac{N^\phi}{N^2} & 0 & \dfrac{-N^2 + r^2(N^\phi)^2}{N^2 r^2} \end{pmatrix}. \qquad (3.44)$$

Having represented the metric tensor results, it is now easy to come by the determinant of $g_{\mu\nu}$.

$$\det g = \frac{N^2 r^2 - r^4 (N^\phi)^2}{N^2} + \frac{r^4 (N^\phi)^2}{N^2} = r^2$$

(3.45)

We have defined all the expressions we need, thus it is time for deliberating Dirac equation in detail. Substituting $\mu = \{0,1,2\}$ into Eq. (3.41) leads to

$$i\left[\sigma_{DC}^0 (\partial_0 - \Gamma_0) + \sigma_{DC}^1 (\partial_1 - \Gamma_1) + \sigma_{DC}^2 (\partial_2 - \Gamma_2)\right]\Psi_f = m_f \Psi_f, \qquad (3.46)$$

where the only expression we have not defined is the connection term. The reason why we have left the accounting of $\Gamma_\mu$ to the end is because it is the one that is in need of the most attentive dealing. Its definition includes spin operator and Christoffel symbol components. Moreover, the ordinary derivatives of tetrads should be computed. It is more harmonious to approach the evaluation of spin connection $\Gamma_\mu$ step by step.

### 3.3.1 Estimation of Spin Affine Connection $\Gamma_\mu$

- Step 1 - Definition of Spin Connection

One may define the connection term as (Sucu & Unal, 2007)

$$\Gamma_\mu = \Pi_{\alpha\mu\gamma} s^{\alpha\mu}, \qquad (3.47)$$

where $s^{\alpha\mu}$ represents the spin operator and the tensor $\Pi_{\alpha\mu\gamma}$ can be further illustrated as

$$\Pi_{\alpha\mu\gamma} = \frac{1}{4} g_{\alpha\beta} \left(e^k_{\mu,\gamma} e^\beta_k - \Gamma^\beta_{\mu\gamma}\right). \qquad (3.48)$$

- Step 2 - Definition of Vierbeins

In this step, we will be checking the metric (3.42) to specify the 'vierbeins' to be considered. By definition (Gecim & Sucu, 2013),

$$g_{\mu\nu} = e_\mu^m e_\nu^k \eta_{mk}, \tag{3.49}$$

where $\eta_{mk}$ is the Minkowski metric tensor and in our case the metric signature is $(+,-,-,-)$.

- Step 3 - Vierbein Computation

The only non-zero components of covariant metric tensor (3.43) are the diagonal terms together with $g_{02}$ and $g_{20}$. Therefore, as expected from its name, only four vierbeins survive and the methodology of obtaining them is demonstrated below.

According to Eq. (3.49), one reads

$$g_{00} = e_0^m e_0^k \eta_{mk} = e_0^0 e_0^0 \eta_{00}, \tag{3.50}$$

$$g_{11} = e_1^m e_1^k \eta_{mk} = e_1^1 e_1^1 \eta_{11}, \tag{3.51}$$

$$g_{22} = e_2^m e_2^k \eta_{mk} = e_2^2 e_2^2 \eta_{22} \tag{3.52}$$

and

$$g_{02} = g_{20} = e_0^m e_2^k \eta_{mk} = e_0^0 e_2^0 \eta_{00} + e_0^2 e_2^2 \eta_{22}. \tag{3.53}$$

Notice that even though *m* and *k* are dummy indices, I have not written all the components from 0 to 2 explicitly. This is due to the simplification that I have carried out, only the non-zero terms will survive, so there is no need to write down each of them one by one. Consequently, the tetrads can be expressed in a table designed below.

Table 2. The vierbein component values.

| The tetrad component symbol | Value of the concerned tetrad |
|---|---|
| $e_t^0$ | $N$ |
| $e_r^1$ | $\dfrac{1}{N}$ |
| $e_\phi^2$ | $r$ |
| $e_\phi^0$ | $-\dfrac{N}{N^\phi}$ |

These results can be plugged into Table 1 and the Dirac matrices in curved manifold will take the form

$$\sigma_{DC}^\mu = \left( -\frac{\sigma_P^3}{N}, -\mathrm{i} N \sigma_P^1, \frac{rN^\phi \sigma_P^3 - iN\sigma_P^2}{Nr} \right). \tag{3.54}$$

- Step 4 - Calculating $\Pi_{\alpha\mu\gamma}$ Tensor

The tetrad results illustrated in Table 2 will now be used to calculate $\Pi_{\alpha\mu\gamma}$ tensor. Recalling Eq. (3.48) and substituting the concerned vierbeins together with their derivatives, the covariant metric tensor and Christoffel symbols, one obtains the non-zero components of $\Pi_{\alpha\mu\gamma}$ as

$$\Pi_{100} = -\Pi_{010} = \frac{1}{4}\left[ \frac{1}{2}\left(N^2\right)' - r\left(N^\phi\right)^2 - r^2 N^\phi \left(N^\phi\right)' \right], \tag{3.55}$$

$$\Pi_{212} = -\Pi_{122} = \frac{r}{4}, \tag{3.56}$$

$$\Pi_{012} = \Pi_{210} = -\Pi_{120} = -\Pi_{102} = \frac{1}{4}\left[ \frac{1}{2}r^2\left(N^\phi\right)' + rN^\phi \right] \tag{3.57}$$

and

$$\Pi_{021} = -\Pi_{201} = \frac{1}{8}r^2\left(N^\phi\right)'. \qquad (3.58)$$

- Step 5 - Evaluation of Spin Operators

By using (Sakurai & Napolitano, 2014)

$$s^{\alpha\mu} = \frac{1}{2}\left[\sigma_{DC}^\alpha, \sigma_{DC}^\mu\right] \qquad (3.59)$$

for the spin operator calculation process, the spin affine connection terms specified in Eq. (3.47) finally become

$$\Gamma_0 = \Pi_{\alpha 0\gamma}s^{\alpha 0} = \frac{1}{4}\left[(N^2)' - r^2 N^\phi \left(N^\phi\right)'\right]\sigma_P^2 + \frac{iN}{4}\left[2N^\phi + r\left(N^\phi\right)'\right]\sigma_P^3, \qquad (3.60)$$

$$\Gamma_1 = \Pi_{\alpha 1\gamma}s^{\alpha 1} = \frac{r\left(N^\phi\right)'}{4N}\sigma_P^1 \qquad (3.61)$$

and

$$\Gamma_2 = \Pi_{\alpha 2\gamma}s^{\alpha 2} = -\frac{r^2\left(N^\phi\right)'}{4}\sigma_P^2 + \frac{iN}{2}\sigma_P^3. \qquad (3.62)$$

### 3.3.2 Further Investigation on DE

Prosperously, we have managed to construe all the components appearing in DE. Now comes the final stage in which the appropriate substitutions should be made into Eq. (3.46). Plugging Eqns. (3.54), (3.60), (3.61) and (3.62) into Eq. (3.46) results in

$$-i\frac{\sigma_P^3}{N}\partial_0\Psi_f + N\sigma_P^1\partial_1\Psi_f + \left(\frac{1}{r}\sigma_P^2 + i\frac{N^\phi}{N}\sigma_P^3\right)\partial_3\Psi_f + \left(\frac{(N^2)'}{4N} + \frac{N}{2r}\right)\sigma_P^1\Psi_f - \frac{\left(N^\phi\right)'r}{4}\sigma_P^0\Psi_f = m_f\Psi_f \qquad (3.63)$$

where $\sigma_P^0$ symbolizes the identity matrix. It is quite clear that we need an ansatz for the wave function of the fermions $\Psi_f$. Let us assume that

$$\Psi_f = \begin{pmatrix} \varphi_1(t,r,\phi)\exp\left[\dfrac{i}{\hbar}I(t,r,\phi)\right] \\ \varphi_2(t,r,\phi)\exp\left[\dfrac{i}{\hbar}I(t,r,\phi)\right] \end{pmatrix} \tag{3.64}$$

in which the functions $\varphi_1(t,r,\phi)$ and $\varphi_2(t,r,\phi)$ can be determined by using the coefficient matrix method. In this method, we generate a matrix whose determinant is set to be zero. By this way, one can successfully obtain two coupled equations which will consequently be used to evaluate the radial function. The coefficient matrix method will be explained in detail throughout Section 4.2.1, hence for now let us express the equation obtained via the application of this method.

$$\frac{1}{N^2}\left(\partial_0 I - N^\phi \partial_2 I\right)^2 - N^2\left(\partial_1 I\right)^2 - \frac{1}{r^2}\left(\partial_2 I\right)^2 - m_f^2 = 0 \tag{3.65}$$

where the radial action can be derived from

$$\omega_\pm(r) = \pm\int \frac{\sqrt{\tilde{E}^2 - N^2\left[\left(\dfrac{h}{r}\right)^2 + m_f^2\right]}}{N}, \tag{3.66}$$

where

$$\tilde{E} = E + hN^\phi. \tag{3.67}$$

Finally, the HT of our RHSBH can be written as

$$T_H = \frac{(B+r_+)\left[36r_+ - J^2\ell^2(2B+3r_+)^2\right]\hbar}{24\pi\ell^2 r_+^5(2B+3r_+)}. \tag{3.68}$$

Note that the same steps as in section 2.2.2 should be followed, with an additional function $N^\phi$.

# Chapter 4

# HR OF THE VECTOR PARTICLES IN THE WAdS$_3$BH GEOMETRY

## 4.1 The Comparison of Ordinary and Warped Anti-de Sitter BHs

Prior to carrying out a deep investigation on WAdS$_3$BHs, one should first ask himself the question of why it is of great importance. As a matter of fact, Minkowski spacetime experiences a zero curvature (Aldrovandi, Almeida, & Pereira, 2007). In the case when the cosmological constant exists, we can no longer consider a zero curvature and this gives rise to the necessity of introducing dS / adS geometries. Moschella (Moschella, 2006) designated dS geometry as "the role of reference geometry of the universe if one describes dark energy with cosmological constant". To have a better understanding of dS space, one can visualize the case of having a hyperboloid with one sheet. The line element of AdS$_n$ embedded in (n+1) dimensions can be expressed as (Moschella, 2006)

$$ds^2 = (dx_0)^2 - (dx_1)^2 + \ldots + (dx_{n-1})^2. \qquad (4.1)$$

There exists a constant curvature and a cosmological constant which are both turned out to have negative values. By working in AdS space, we can also overcome the logical inconsistency in the compulsion of placing a BH in a box that has a finite heat capacity to evaluate the thermodynamically stable solutions (Brito, Cardoso, & Pani, 2015). It is also worthwhile to record that CFT, ADM formalism and AdS/CFT Correspondence play a vital role in the investigation of the BH family belonging to AdS$_3$ space.

As might have been noticed from the title of this section, BHs in AdS space can be categorized into two groups; ordinary and warped $AdS_n$ BHs. In this thesis, I will be dealing with two space coordinates, namely $r$ and $\varphi$, and time coordinate $t$, and hence $AdS_3$ geometry will be considered.

You might be wondering how a $WAdS_3BH$ differentiates itself from a BTZ BH. Knowing that both of these mysterious objects are members of the AdS space, what is the main difference we should seek for? To be able to answer this question, let us first have a brief look at the geometrical structure of BTZ BHs. The line element of BTZ BHs can be constructed as (Li, Li, & Ren, 2011)

$$ds^2 = -N_{BTZ}^2 dt^2 + \frac{1}{N_{BTZ}^2} dr^2 + r^2 \left(d\phi + N_{BTZ}^\phi dt\right)^2, \qquad (4.2)$$

where

$$N_{BTZ}^2 = -M + \frac{r^2}{\ell^2} + \frac{J}{4r^2} \qquad (4.3)$$

and

$$N_{BTZ}^\phi = \frac{-J}{2r^2}.$$

(4.4)

BTZ and $WAdS_3$ BHs differ in their structure and for the warped case; the symmetry is trimmed down (Birmingham & Mokhtari, 2011). A $WAdS_3BH$ is characterized as (Gursel & Sakalli, 2015)

$$ds^2 = -N^2 dt^2 + \frac{\ell^4}{4R^2 N^2} dr^2 + \ell^2 R^2 \left(d\phi + N^\phi dt\right)^2, \qquad (4.5)$$

where

$$R^2 = \frac{r}{4}\left[3(\upsilon^2-1)r+(\upsilon^2+3)(r_++r_-)-4\upsilon\sqrt{r_+r_-(\upsilon^2+3)}\right], \quad (4.6)$$

$$N^2 = \frac{(\upsilon^2+3)(r-r_+)(r-r_-)}{4R(r)^2} \quad (4.7)$$

and

$$N^\phi = \frac{2\upsilon r - \sqrt{r_+r_-(\upsilon^2+3)}}{2R(r)^2}. \quad (4.8)$$

The functions $N^\phi(r)$, $N(r)^2$ and $R(r)$ include a constant $\upsilon$ of which value being equal to 1/3 implies chiral gravity theory (Anninos, Li, Padi, Song, & Strominger, 2009). By excluding this chiral point, we are left with the possible BH solutions arising, depending on the value that $\upsilon$ has. These can be tabulated as follows.

Table 3. Classification of WAdS3BHs

| Value of $\upsilon$ | BH Solution |
|---|---|
| < 1 | *Squashed AdS$_3$* |
| > 1 | *Streched AdS$_3$* |
| = 1 | *Null AdS$_3$* |

Table 3 is composed of three elements, however it should be noted that each of the specified black hole solution can be separated into two sub groups as spacelike and timelike. Thus, it consequently follows that six solutions are present in total.

## 4.2 QT of Massive Vector Particles from Spacelike Streched WAdS₃BHs using PEs

In compact notation, PE can be written as (Kruglov, 2014)

$$D_\mu \Psi^{\lambda\mu} + \frac{m^2}{\hbar^2}\Psi^\lambda = 0, \tag{4.9}$$

where

$$\Psi_{\lambda\mu} = D_\lambda \Psi_\mu - D_\mu \Psi_\lambda = \partial_\lambda \Psi_\mu - \partial_\mu \Psi_\lambda. \tag{4.10}$$

It is clear from our metric (4.5) that $\mu = \{0,1,2\}$ and since the index $\mu$ is repeated in Eq. (4.9), the three components should be summed up for each $\lambda$ value as stated below.

$$D_0 \Psi^{\lambda 0} + D_1 \Psi^{\lambda 1} + D_2 \Psi^{\lambda 2} + \frac{m^2}{\hbar^2}\Psi^\lambda = 0. \tag{4.11}$$

It is prudent to remark that the indices should be lowered, since we are keen to use definition (4.10). After a set of calculations, we consequently obtained a set of three equations which can be written explicitly as follows.

$$m^2 \ell^2 R^2 \left(\Psi_t - N^\phi \Psi_\phi\right) + \hbar^2 \ell^2 \partial_\phi \left(\Psi_{t\phi}\right) + 4\hbar^2 R^2 N^2 \partial_r \left(R^2 \frac{\partial \Psi_{tr}}{\partial t} + N^2 \frac{\partial \Psi_{r\phi}}{\partial \phi}\right) = 0,$$

$$m^2 R^2 N^2 \Psi_r + \hbar^2 \left(N^\phi\right)^2 R^2 \left(\frac{\partial \Psi_{r\phi}}{\partial t} - \frac{\partial}{\partial \phi}\left[\Psi_{tr} + \Psi_{r\phi}\right]\right) + \hbar^2 \left(R^2 \frac{\partial \Psi_{tr}}{\partial t} + N^2 \frac{\partial \Psi_{r\phi}}{\partial \phi}\right) = 0$$

and

$$m^2 \ell^2 R^2 \left(\Psi_t - N^\phi \Psi_\phi\right) + \hbar^2 \ell^2 \partial_\phi \left(\Psi_{t\phi}\right) + 4\hbar^2 R^2 N^2 \left(\partial_r \left[R^2 \Psi_{tr}\right] + \partial_r \left[N^\phi R^2 \Psi_{r\phi}\right]\right) = 0 \tag{4.12}$$

### 4.2.1 Obtaining the Radial Function $\omega_{\pm}(r)$ by Coefficient Matrix Method

Having evaluated the PEs, one shall now assume an ansätz for the wavefunction $\Psi_\lambda$ in the exponential form and investigate the equations with the aim of radial function assessment. Let the wave function be expressed as

$$\Psi_\lambda = c_\lambda e^{\frac{iI(t,r,\phi)}{\hbar}}, \tag{4.13}$$

where $I(t,r,\varphi)$ and $c_\lambda$ are the action and coefficient respectively. There are many different varieties of approaches that we can derive a profit from to solve PEs, however the most practical one is presumably the coefficient matrix method. In this method, the coefficients are collected in a matrix, namely $\mathbf{M}$, which is summarized in the Table 4.

Table 4. Components of the Coefficient Matrix.

| Components of $M_{ab}$ | Value of Component |
|---|---|
| $M_{00}$ | $-4R^4 N^2 \omega'^2 - \ell^2 \left( m^2 R^2 + h^2 \right)$ |
| $M_{01} = M_{10}$ | $-4R^4 N^2 \omega' \left( E + hN^\phi \right)$ |
| $M_{02} = M_{20}$ | $-\ell^2 hE + m^2 \ell^2 R^2 N^\phi + 4R^4 N^2 N^\phi \omega'^2$ |
| $M_{12} = M_{21}$ | $4\omega' \left( NR^2 \left[ E + hN^\phi \right] - N^2 h \right) N^2 R^2$ |
| $M_{11}$ | $-4N^2 R^2 \left( \left[ -m^2 N^2 + \{E + hN^\phi\}^2 \right] R^2 - N^2 h^2 \right)$ |
| $M_{22}$ | $\left( -4R^4 N^2 (N^\phi)^2 + 4R^2 N^4 \right) \omega'^2 - \ell^2 \left( E^2 + m^2 R^2 (N^\phi)^2 - m^2 N^2 \right)$ |

The next step is to find the determinant of the matrix $M$, equate it to zero and solve for the radial function $\omega(r)$. The determinant of any $3x3$ matrix can be determined by performing the fairly straightforward routine stated below. If

$$A = \begin{pmatrix} A_{11} & A_{12} & A_{13} \\ A_{21} & A_{22} & A_{23} \\ A_{31} & A_{32} & A_{33} \end{pmatrix}, \tag{4.14}$$

then the determinant reads (Arfken, 2013)

$$\det(A) = |A| = A_{11}\begin{vmatrix} A_{22} & A_{23} \\ A_{32} & A_{33} \end{vmatrix} - A_{12}\begin{vmatrix} A_{21} & A_{23} \\ A_{31} & A_{33} \end{vmatrix} + A_{13}\begin{vmatrix} A_{21} & A_{22} \\ A_{31} & A_{32} \end{vmatrix}. \tag{4.15}$$

Hence, applying the same procedure to the matrix $M$ leads to

$$\det(M) = -4R^2N^2m^2\left(-4R^4N^4\omega'^2 + \left[\left\{-m^2N^2 + \left(E+hN^\phi\right)^2\right\}R^2 - N^2h^2\right]\ell^2\right)^2 \tag{4.16}$$

As had been previously mentioned, setting $\det(M) = 0$ is a compulsory requirement of the progression. By doing so and applying WKB approximation on the action, we only need to take

$$I_0 = -Et + h\phi + \omega(r) + \aleph, \tag{4.17}$$

into account, same as Eq. (2.11). We finally have achieved an appropriate expression for the radial function $\omega(r)$:

$$\omega_{\pm}(r) = \pm\int dr \left(\frac{\ell}{2RN^2}\right)\left(\sqrt{\left[E+hN^\phi\right]^2 - N^2\left[m^2 + \frac{h^2}{R^2}\right]}\right)^{1/2} \tag{4.18}$$

The radial function attained is hiding two solutions in itself; one for the ingoing and another for the outgoing particles. This will almost certainly involve further analysis in order for determining which one will be of use throughout the HT evaluation process.

### 4.2.2 HT Calculation on the Event Horizon

The vacuum fluctuations are believed to be taking place at the event horizon limit of the black hole, or in other words, during our calculations, it will be appropriate to consider $r \to r_+$. This will promptly affect the metric function $N(r)$, making it approach to zero. Subsequently, the radial function expression will reduce to

$$\omega_\pm(r) = \pm \frac{\ell}{2} \int \frac{E + hN^\phi}{RN^2} dr. \qquad (4.19)$$

Whilst an integral of the form (4.19) is faced, it is usually convenient to discuss the calculus of residues. To be able to do so, one should first substitute the metric functions in Eq. (4.20) and this results in

$$\omega_\pm(r) = \pm 2\ell \int dr \frac{E + hN^\phi}{(\upsilon^2 + 3)(r - r_+)(r - r_-)} \sqrt{\frac{r}{4}\left(3[\upsilon^2 - 1]r + \Xi\right)} \qquad (4.20)$$

in which

$$\Xi = (\upsilon^2 + 3)(r_+ + r_-) - 4\upsilon\sqrt{r_+ r_-(\upsilon^2 + 3)}. \qquad (4.21)$$

According to complex contour analysis, under the condition that $f(z)$ is an analytic function on our contour, the following relation holds.

$$\frac{1}{\pi i} \int \frac{f(z)dz}{z - z_0} = f(z_0), \qquad (4.22)$$

iff $z_0$ is an interior singular point. In our case, the singular point is $r_+$. So, Eq. (4.23) shall be rewritten as

$$\frac{1}{\pi i} \int \frac{f(r)dr}{r - r_+} = f(r_+)$$

.

(4.23)

Comparing Eq. (4.21) with Eq. (4.24), it can clearly be seen that

$$f(r) = \frac{E + hN^\phi}{(v^2+3)(r-r_-)} \sqrt{\frac{r}{4}\left(3\left[v^2-1\right]r + \Xi\right)}, \qquad (4.24)$$

which modifies the radial function to the form

$$\omega(r_+) \cong i\pi\ell \frac{E + hN^\phi}{(v^2+3)(r_+ - r_-)} \sqrt{3r_+^2(v^2-1) + \Xi r_+} \ . \qquad (4.25)$$

The task of finding the HT can now be accomplished, since the radial function is stated in a compact notation. All in all, the surface temperature of this BH, namely the HT, can be written as

$$T_H = \frac{E}{4\pi\omega_+}. \qquad (4.26)$$

Substituting Eqn. (4.25) into Eqn. (4.26) results in

$$T_H = \frac{(v^2+3)(r_+ - r_-)}{4\pi\ell\left[2vr_+ - \sqrt{(v^2-1)r_+r_-}\right]}. \qquad (4.27)$$

# Chapter 5

# CONCLUSION

This thesis utilizes various techniques of evaluating the invariant HT expression persisting unaltered for different types of BHs. Virtual pair (particle − antiparticle) creation and QT were the processes on which these evaluations were based and in this manner, it is essential to bear in mind that each particle being tunneled from the event horizon of our BH might possibly have a different spin value. We were obliged to seek whether this plays a role in the HT expression or not. To pursue this problem, we have divided the thesis into different chapters in which spin-0, spin-1/2 and spin-1 particles were investigated individually; and in the end we have shown that in all cases, one unique expression has been attained: $T_H = \dfrac{\kappa \hbar}{2\pi}$. The result obtained was in harmony with the HT derived in Hawking's original work (S. Hawking, 1975). Consistent with previous research, the results suggest that there is no linkage between the spin of the particle being emitted from a BH and the HT acquired.

Although studies have discussed the importance of evaluation of HT via using QT process, a challenge that has received relatively little attention is for WAdS$_3$BHs and RHSBHs, as only few have actually focused on the HR being radiated from these BHs specifically. The core theoretical contribution of chapter 4 is the suggestion that PEs can actually be used in the evaluation of HT of a BH when vector particles are taken into account. Furthermore, chapter 3 provided us with more knowledge regarding under what circumstances no hair theorem can be deserted.

For future studies, the effect of magnetic monopoles on HR of BHs can be investigated, which seems to be an inspiring topic; since the magnetic charge could actually be a part of the no hair theorem and it can lead to conceivable contributions in QM like equipping us with an explanation of why charges are found as quantized in nature. One can turn these new ideas over her/his mind, attempting to hit upon some theory to catch that line of least resistance, which possibly is the starting point of every discovery.

I would like to finish my thesis with the words of Morgan Freeman: "Gravity feels real. But, gravity may not be what it seems. If gravity is an illusion, then it is time to call into the question everything we think we know about the cosmos. Only if we let go of what we feel to be correct; can we taste the real."

# REFERENCES

Akhmedov, E. T., Akhmedova, V., & Singleton, D. (2006). Hawking temperature in the tunneling picture. *Physics Letters B, 642*(1), 124-128.

Aldrovandi, R., Almeida, J., & Pereira, J. (2007). Some implications of the cosmological constant to fundamental physics. *arXiv preprint gr-qc/0702065*.

Anninos, D., Li, W., Padi, M., Song, W., & Strominger, A. (2009). Warped AdS3 black holes. *Journal of High Energy Physics, 2009*(03), 130.

Arfken, G. B. (2013). *Mathematical methods for physicists*: Academic press.

Bardeen, J. M., Carter, B., & Hawking, S. W. (1973). The four laws of black hole mechanics. *Communications in mathematical physics, 31*(2), 161-170.

Bekenstein, J. D. (1973). Black holes and entropy. *Phys. Rev. D7*.

Bekenstein, J. D. (1995). Novel ''no-scalar-hair''theorem for black holes. *Physical Review D, 51*(12), R6608.

Birmingham, D., & Mokhtari, S. (2011). Thermodynamic stability of warped AdS3 black holes. *Physics Letters B, 697*(1), 80-84.

Bojowald, M. (2012). *The Universe: A View from Classical and Quantum Gravity*: John Wiley & Sons.

Brito, R., Cardoso, V., & Pani, P. (2015). Black Holes and Superradiant Instabilities *Superradiance* (pp. 97-155): Springer.

Carroll, S. M. (2001). A No-Nonsense Introduction to General Relativity. *Enrico Fermi Institute and Department of Physics, University of Chicago*.

Carroll, S. M. (2004). *Spacetime and geometry. An introduction to general relativity* (Vol. 1).

Condon, E. (1929). u. PM Morse: Quantum Mechanics: New York: McGraw-Hill.

Cox, D. A. (2011). *Galois theory* (Vol. 61): John Wiley & Sons.

Dirac, P. A. M. (1981). *The principles of quantum mechanics*: Oxford university press.

Einstein, A. (1928a). New possibility for a unified field theory of gravitation and electricity. *Session Report of the Prussian Acad. Sci*, 224-227.

Einstein, A. (1928b). Riemannian Geometry with Maintaining the Notion of distant parallelism. *Session Report of the Prussian Academy of Sciences*, 217-221.

Einstein, A. (1952). The foundation of the general theory of relativity. *The Principle of Relativity. Dover Books on Physics. June 1, 1952. 240 pages. 0486600815, p. 109-164, 1*, 109-164.

Falls, K., & Litim, D. F. (2014). Black hole thermodynamics under the microscope. *Physical Review D, 89*(8), 084002.

Gecim, G., & Sucu, Y. (2013). Tunnelling of relativistic particles from new type black hole in new massive gravity. *Journal of Cosmology and Astroparticle Physics, 2013*(02), 023.

Grössing, G. (2002). Derivation of the Schroedinger Equation and the Klein-Gordon Equation from First Principles. *arXiv preprint quant-ph/0205047*.

Gursel, H., & Sakalli, I. (2015). Hawking Radiation of Massive Vector Particles From Warped AdS $ _ {\ text {3}} $ Black Hole. *arXiv preprint arXiv:1506.00390*.

Hartle, J. B. (2002). *Gravity: an introduction to Einstein's general relativity*: Pearson Education India.

Hawking, S. (1975). Commun.,". *Particle Creation By Black Holes", Math. Phys, 43*, 199.

Hawking, S. W. (1975). Particle creation by black holes. *Communications in mathematical physics, 43*(3), 199-220.

Hobson, M. P., Efstathiou, G. P., & Lasenby, A. N. (2006). *General relativity: an introduction for physicists*: Cambridge University Press.

John D. Cutnell, K. W. J. (2013). *Introduction to Physics* (Vol. 9th): Wiley.

Kaku, M. (1993). *Quantum field theory*: Oxford Univ. Press.

Kruglov, S. (2014). Black hole emission of vector particles in (1+ 1) dimensions. *International Journal of Modern Physics A, 29*(22), 1450118.

Li, R., Li, M.-F., & Ren, J.-R. (2011). Hidden conformal symmetry of self-dual warped AdS3 black holes in topological massive gravity. *The European Physical Journal C, 71*(2), 1-8.

Mei, T. (2006). On isotropic metric of Schwarzschild solution of Einstein equation. *arXiv preprint gr-qc/0610112*.

Moschella, U. (2006). *The de Sitter and anti-de Sitter sightseeing tour*: Springer.

Resnick, R., Halliday, D., & Walker, J. (1988). *Fundamentals of physics*: John Wiley.

Sakalli, I. (2012). Effect of the cosmological constant in the Hawking radiation of 3D charged dilaton black hole. *Astrophysics and Space Science, 340*(2), 317-322.

Sakalli, I., & Mirekhtiary, S. (2013). Effect of the refractive index on the hawking temperature: an application of the Hamilton-Jacobi method. *Journal of Experimental and Theoretical Physics, 117*(4), 656-663.

Sakalli, I., & Ovgun, A. (2015). Uninformed Hawking radiation. *EPL (Europhysics Letters), 110*(1), 10008.

Sakurai, J. J., & Napolitano, J. J. (2014). *Modern quantum mechanics*: Pearson Higher Ed.

Schroeder, D. V. (2000). Thermal physics: Addison Wesley Longman, San Francisco.

Shankaranarayanan, S., Srinivasan, K., & Padmanabhan, T. (2001). Method of complex paths and general covariance of Hawking radiation. *Modern Physics Letters A, 16*(09), 571-578.

Stephani, H., Kramer, D., MacCallum, M., Hoenselaers, C., & Herlt, E. (2003). *Exact solutions of Einstein's field equations*: Cambridge University Press.

Strorninger, A., & Vafa, C. (1996). Microscopic Origin of the Bekenstein-Hawking Entorpy. *arXiv preprint hep-th/9601029*.

Sucu, Y., & Unal, N. (2007). Exact solution of Dirac equation in 2+ 1 dimensional gravity. *Journal of mathematical physics, 48*(5), 52503-52700.

Toubal, W. (2010). *No-hair theorems and introduction to Hairy black holes* (Master of Science ), Imperial College London

Young, H. D., & Freedman, R. A. (2008). University Physics, Vol. I: Pearson, Addison Wesley.

Zou, D.-C., Liu, Y., Wang, B., & Xu, W. (2014). Rotating black holes with scalar hair in three dimensions. *arXiv preprint arXiv:1408.2419*.

Zou, D.-C., Liu, Y., Zhang, C.-Y., & Wang, B. (2014). Dynamical probe of thermodynamical properties in three-dimensional hairy AdS black holes. *arXiv preprint arXiv:1411.6740*.